\documentclass[12pt,thmsa]{article}

\input{tcilatex}
\begin{document}

\author{M. Tierz \\
Institut d'Estudis Espacials de Catalunya \\
(IEEC/CSIC), Edifici Nexus, Gran \\
Capit\`{a}, 2-4, 08034 Barcelona, Spain. \\
tierz@ieec.fcr.es}
\date{}
\title{The Stable Random Matrix ensembles}
\maketitle

\begin{abstract}
We address the construction of stable random matrix ensembles as the
generalization of the stable random variables (Levy distributions). With a
simple method we derive the Cauchy case, which is known to have remarkable
properties. These properties allow for such an intuitive method -that relies
on taking traces- to hold. Approximate but general results regarding the
other distributions are derived as well. Some of the special properties of
these ensembles are evidenced by showing partial failure of mean-field
approaches. To conclude, we compute the confining potential that gives a
Gaussian density of states in the limit of large matrices. The result is an
hypergeometric function, in contrast with the simplicity of the Cauchy case.
\end{abstract}

\section{Introduction}

The ensembles of random matrices have shown beyond doubt its usefulness in
many physical applications \cite{Meh}. The definition of the joint
probability distribution $P\left( M\right) $ of the matrix elements of a $N$
by $N$ matrix $M$ is \cite{Meh}: 
\begin{equation}
P(M)=C_{N}\exp [-Tr(V(M))]~,  \label{eq1}
\end{equation}
with an arbitrary $V\left( M\right) ,$ provided existence of the partition
function $C_{N}^{-1}$. Integrating $\left( \ref{eq1}\right) $ over the
parameters related to the eigenvectors, one obtains the well-known joint
probability distribution of the eigenvalues: 
\begin{equation}
P(x_{1},...,x_{N})=C_{N}\exp [-\sum_{i=1}^{N}V(x_{i})]\prod_{i<j}\left|
x_{i}-x_{j}\right| ^{\beta }~.
\end{equation}

Level repulsion described by the Vandermonde determinant is originated from
the Jacobian, that appears when passing from the integration over
independent elements of the Hamiltonian to the integration over the smaller
space of its $N$ eigenvalues. The parameter $\beta ,$ with values $1,2$ or $%
4 $ describe the symmetry of the ensemble (named orthogonal, unitary and
symplectic, respectively).

If the elements of the random matrix are believed to be statistically
independent from each other, one obtains the quadratic confinement
potential, leading to the Gaussian ensembles of random matrices \cite{Meh}.
The requirement of statistical independence is not motivated from first
principles. Thus, the influence of a particular form for the confinement
potential in relationship with the Gaussian predictions is an interesting
problem \cite{Kan}.

As it is well-known, the joint probability distribution can be mapped onto
the Gibbs distribution of a classical one-dimensional gas of fictitious
particles with a pair-wise logarithmic repulsion and a one-particle
potential.

\begin{equation}
P(\{x_{i}\})=Z^{-1}\exp[-\beta H(\{x_{i}\})]~,
\end{equation}

\begin{equation}
H(\{x_{i}\})=-\sum_{i<j}\ln \left| x_{i}-x_{j}\right| +\sum_{i}V(x_{i})~.
\end{equation}
This is usually called the Coulomb gas picture. Let us now present the
mean-field approximation that was introduced by Dyson \cite{Dys}. By
substituting the density of states in the previous Hamiltonian, we obtain a
continuous description of the energy functional $H[\rho ]$ in terms of $\rho
(x)$. The extremum of this functional corresponds to an equilibrium of the
effective plasma expressed by the equation: 
\begin{equation}
\int dx^{\prime }\rho (x^{\prime })ln\left| x-x^{\prime }\right| =V(x)+c~,
\end{equation}
where $\rho $ is the mean density. This expression is usually known as
Wigner equation. In \cite{Dys}, the next correction term was found using a
hydrodynamical approximation:

\begin{equation}
V(x^{\prime })=\int_{-\infty }^{\infty }\rho (x)\ln \left| x-x^{\prime
}\right| dx+\frac{1}{2}(1-\frac{\beta }{2})\ln \rho (x^{\prime })~.
\label{Dyseq}
\end{equation}

Higher order terms involve a complicate combination of the density of levels
with correlation functions \cite{Been2}. In addition, in {\cite{Dys} it} was
shown that in this approximation the first correction term -the one that
appears in $\left( \ref{Dyseq}\right) $- is of order $\frac{\ln N}{N}$,
while the next term is of order $N^{-2}$. This applies to systems with
strong confining potential.

In this paper, we study features of stable random matrix ensembles, in
analogy with the stable random variables. We provide here a derivation,
based on simple arguments, of the Cauchy case for the three ensembles ($%
\beta =1,2$ and $4$). In this way, we obtain the so-called Lorentzian
ensemble, that has been already studied with various points of view {\cite
{Been2,Brou,Hua,Witt}}. Further work related with stable distributions can
be found in \cite{Wie}, and in \cite{Berc} from the point of view of free
probability theory \cite{Voi}.

In section 3, this ensemble allows to provide explicit examples to the
failure of Dyson equation, in the case of non strongly confining potentials.
Additionally, we solve several inverse problems using Dyson equation and
present a systematic comparison between all cases. In particular, we find
the confining potential that gives a Gaussian density of states in the limit 
$N\rightarrow \infty $.

\section{The stable random matrix ensembles. Derivation}

We begin by showing a simple way to obtain the expression for the stable
ensembles. We will focus on the Cauchy distribution, but all the stable
distributions can be studied, in principle, with the same method. We look
for ensembles that satisfy the following: 
\begin{equation}
M=\sum_{i=1}^{N}M_{i}~.
\end{equation}
$M$ and $M_{i}$ denote the random matrices on the same ensemble. Put it
simply, the matrix generalization of the stable random variables (see
Appendix A). Notice the following: 
\begin{equation}
\mathrm{Tr}M=\mathrm{Tr}\sum_{i=1}^{N}M_{i}=\sum_{i=1}^{N}\mathrm{Tr}M_{i}~.
\label{trace}
\end{equation}
By taking the trace of the matrix, we reduce the problem to a problem of
random variables. In order that this last equality holds, it is evident that
the trace itself should be a stable random variable. More clearly, we make a
mapping between the stable random variables and the stable matrices (the
matrix counterpart). It is manifest now in which sense we say that we search
for the matrix generalization of the stable random variables. Note that the
expression for the Gaussian ensemble satisfy the relation $\left( \ref{trace}%
\right) $, after taking the trace on its $P(x_{1},...,x_{N})$. Let us go now
into the Cauchy case. We begin by writing again the general expression in
terms of the eigenvalues:

\begin{equation}
P(x_{1},...,x_{N})=C_{N}\prod_{i<j}\left| x_{i}-x_{j}\right| ^{\beta }\exp
[-\sum_{i=1}^{N}V(x_{i})]~.
\end{equation}
For consistency with the case $N=1$, we clearly need the following weight
function (or confining potential): 
\begin{equation}
\omega (x)=\exp (-V(x))=\frac{1}{\lambda ^{2}+x^{2}}~.
\end{equation}

It is also clear that we need some function of $N$ in our weight function.
Otherwise, the joint probability distribution would not even be
normalizable. What we need to impose is: 
\begin{equation}
\int_{-\infty }^{\infty }\cdot \cdot \cdot \int_{-\infty }^{\infty
}dx_{1}\cdot \cdot \cdot dx_{N}P(x_{1},...,x_{N})\delta
(y-x_{1}-x_{2}-...-x_{N})=\frac{\lambda N}{\pi }\frac{1}{(\lambda
N)^{2}+y^{2}}~,
\end{equation}
with: 
\begin{equation}
P(x_{1},...,x_{N})=C_{N}\prod_{i<j}\left| x_{i}-x_{j}\right| ^{\beta
}\prod_{k=1}^{N}\left( \frac{1}{\lambda ^{2}+x^{2}}\right) ^{\alpha (N)}~.
\end{equation}
From the case $N=1$, we already know that $\alpha (1)=1$. We can see that
the function $\alpha (N)$ should be a linear polynomial in $N$. Notice that
we are studying the modification we have to impose to the Cauchy
distribution in order that the effect introduced by the correlation term
does not modify the stability property. The Vandermonde determinant
introduces the product of $N(N-1)/2$ polynomial factors, while the weight
function term introduces $N$ products of polynomial factors. Thus, we need
each one of these factors to contain a term of order $N$. So, we have to
consider the following function:

\begin{equation}
\alpha (N)=a_{0}N+a_{1}~.
\end{equation}
Then, it is enough to look at the $N=2$ case. Needless to say, it can be
checked that, for example, the case $N=3$ implies a zero coefficient for an
hypothetical quadratic term, and so on. Actually, in the next section we
will present an alternative argument that also implies that the function
should be of the form indicated above.

Doing the integral for the three ensembles, we finally arrive at: 
\begin{equation}
P(x_{1},...,x_{N})=\frac{2^{\frac{\beta }{2}(N^{2}-N)}}{N!\pi ^{N}}%
\prod_{i<j}\left| x_{i}-x_{j}\right| ^{\beta }\prod_{k=1}^{N}\left( \frac{1}{%
\lambda ^{2}+x^{2}}\right) ^{(1+\frac{\beta }{2}(N-1))}~.  \label{Lens}
\end{equation}

We have arrived at a known and important random matrix ensemble \cite
{Brou,Hua}. In the physics literature it is named Lorentzian ensemble \cite
{Brou}. The normalization constant can be computed with the Selberg integral 
\cite{Meh}.

\subsection*{2.1. The relationship with the circular ensemble and
universality.}

It has been shown that the Lorentzian ensemble $\left( \ref{Lens}\right) $
has a density of states with same form for any dimension of the matrix and
for any $\beta $ \cite{Brou}. Recall that the same result holds for the
circular ensembles: they have constant density on the circle for all $N$ 
\cite{Meh}. We know that there are several different parametrizations for a
unitary matrix. A rather convenient specific parametrization is $S=\frac{1-iA%
}{1+iA}$ (Cayley transform), since it is one on one. Notice the effect of
any of this transformation on the Cauchy probability distribution:

\begin{equation}
f(x)=\frac{1}{\pi }\frac{1}{1+x^{2}}\;\text{and}\;x=\mathrm{cotg}(\theta /2),
\end{equation}
then: 
\begin{equation}
f(\theta )=\mid \frac{dx}{d\theta }\mid \frac{1}{\pi (1+\mathrm{cotg}(\theta
/2)^{2})}=\mid \frac{-1}{2\cos ^{2}\theta }\mid \frac{\cos ^{2}\theta }{\pi }%
=\frac{1}{2\pi }.
\end{equation}
$.$

That is, the Cauchy distribution transforms into a uniform probability in
the circle. In \cite{Brou,Hua}, it is shown that under the Cayley Transform,
the Circular ensemble maps into the Lorentzian ensemble. Consequently,
taking into account the universality property for the correlation functions
of the circular ensembles in the limit $N\rightarrow \infty $ \cite{Meh},
the Lorentzian ensemble also exhibits universal behavior in this limit. That
is to say, all the correlation functions of the Lorentzian ensemble behave
as the ones from the Gaussian ensemble in the limit of large matrices \cite
{Brou}.

Additionally, notice that if $X$ is a random variable with probability
distribution function $f(x)=\frac{\lambda }{\pi (\lambda ^{2}+x^{2})}$, then
the random variable $Y=X^{-1}$ has the probability distribution $g(y)=\frac{%
\lambda }{\pi (\lambda ^{2}+x^{2})}$. Namely, the inverse of a Cauchy
distribution is a Cauchy distribution. Again, we see that the same property
holds for the matrix generalization, and in \cite{Hua} it is shown that if $%
M $ is a matrix in the Lorentzian ensemble, then $M^{-1}$ is also a matrix
in the Lorentzian ensemble. This result readily follows by the corresponding
change of variables in the multidimensional probability distribution
function for the eigenvalues.

Therefore, a matrix in the Lorentzian ensemble shares many properties with a
diagonal matrix, whose elements are identically distributed Cauchy
distributions. First of all, the diagonal matrix, after trivial integration
over $N-1$ eigenvalues, also posses a Cauchy density of states
(independently of $N$). Secondly, at the level of the density of states it
also has the correspondence with a uniform density. Furthermore, the
diagonal matrix is equal to its inverse as well, as it readily follows from
the property of the Cauchy distribution. Needless to say, the correlation
properties of the eigenvalues are very different. These properties are not
shared by a random matrix in general and make the Lorentzian ensemble very
useful for physical applications \cite{Brou}.

It is also interesting to discuss a little bit more about universality. As
it is well-known, any strongly confining potential (that is, growing at
least as $V\left( x\right) \sim \left| x\right| $ for $x\rightarrow \infty $%
), gives the same Gaussian\ universality class \cite{Me2} (see also \cite
{Been1}). Here, we deal with a special confining potential, that contains
the dimension of the matrix $N$ in a explicit way: 
\begin{equation}
V_{\beta }(x)=(\frac{\beta N+2-\beta }{2})\ln (\lambda ^{2}+x^{2})~.
\end{equation}
Note that the dependence with the variable $x$ is given by a very soft
confining potential. Nevertheless, it is multiplied by a function that is
linear in $N$, the dimension of the matrix. This is important, since any
random matrix universality regime is understood in the limit $N\rightarrow
\infty $. Thus, at the end, it is the term in $N$ the one that achieves the
universal behavior. This clarifies the mechanism that makes this ensemble to
belong to the Gaussian universality class. The consequence is that the
ensemble exhibits very different global properties (the density of states),
while it has the same behavior at the local scale (all the correlation
functions).

It is patent that all the arguments above hold for any stable distribution
(see Appendix A). As an extension, note that this matrix generalization of
the Levy distributions can be expressed, in general, as: 
\begin{equation}
P(x_{1},...,x_{N})=C_{N}\prod_{i<j}\left| x_{i}-x_{j}\right| ^{\beta
}\prod_{k=1}^{N}\left( f_{\alpha ,\beta ^{\prime }}(x)\right)
^{a_{0}N+a_{1}},\quad a_{0}+a_{1}=1~.
\end{equation}

Needless to say, it is consistent that the only stable distribution with
finite variance, the Gaussian distribution, is the only one that has
different form for $\alpha(N)$ (more precisely $a_{0}=0$ and $a_{1}=\beta$)
in the matrix generalization.

We see that the coefficient $a_{0}$ is actually equal to $a_{0}=\frac{\beta 
}{\mu }=\frac{\beta }{\alpha +1}$. Needless to say, as explained above, the
value of the coefficient depends both on the power of the polynomial in the
numerator (the Vandermonde determinant) and the one in the denominator
(weight function). Thus, this particular value seems to be expected for all
the Levy distributions. Since we know the asymptotic behavior of the
distributions for $x$ large (see Appendix A)$:$%
\begin{equation}
f_{\alpha ,\beta ^{\prime }}(x)\sim \frac{A_{\alpha ,\beta ^{\prime }}}{%
\left| x\right| ^{1+\alpha }}\quad \alpha <2~,
\end{equation}
we have the following confining potential for $x$ large: 
\begin{equation}
V(x)=\left( \frac{\beta }{\alpha +1}N+a_{0}\right) \ln \left( 1+x^{1+\alpha
}\right) \approx \left( \beta N+\frac{a_{0}}{1+\alpha }\right) \ln x~,
\end{equation}
and, for $N$ large and $x$ large: 
\begin{equation}
V(x)=\beta N\ln x~.
\end{equation}
This coincides with the result found in \cite{Bur} for the Hermitian case ($%
\beta =2),$ the one considered there. Notice that, at this level of
approximation, the confining potential is the same for any Levy
distribution. Since this is precisely the universality limit, it is expected
that they all share the same universal properties in the sense explained
above.

\section{The Dyson equation}

We use now the previous results to illustrate the failure of Dyson's
mean-field approximation in certain cases. This has been recently discussed
in the literature, and has proved to be an increasingly interesting topic,
as soon as new type of random matrix ensembles, different from the Gaussian
ones, are studied. As mentioned, Dyson equation is \cite{Dys} (see also \cite
{Been2}): 
\begin{equation}
V(x^{\prime })=\int_{-\infty }^{\infty }\rho (x)\ln \left| x-x^{\prime
}\right| dx+\frac{1}{2}(1-\frac{\beta }{2})\ln \rho (x^{\prime })~.
\end{equation}

We would like to study several nontrivial examples that allow to illustrate
the situation. In the progress of doing this, we will also solve some
interesting inverse problem in random matrix theory.

The Cauchy case allow us for a simple and enlightening example. First, we
begin by asking which is the confining potential that give us a Cauchy
density of states in the limit $N$ very large \footnote{%
Note that we already know the answer, is the confining potential in the
previous section.}. This reduces to compute: 
\begin{equation}
V(x)=\frac{\lambda }{\pi }\int_{-\infty }^{\infty }\frac{\ln \left|
x-x^{\prime }\right| }{\lambda ^{2}+x^{^{\prime }2}}dx=\frac{\lambda }{2\pi }%
\ln (\lambda ^{2}+x^{2})=\frac{N}{2}\ln (\lambda ^{2}+x^{2})~.  \label{confm}
\end{equation}

To begin with, we find that for the Hermitian case ($\beta =2$), this result
gives just half of the right value, since the first correction term is
identically zero for $\beta =2$. Thus, we already see that half of the value
would be contained in the rest of the following correction terms and we
should remember that the first one of these corrections is of order $N^{-2}$
in the strongly confining paradigm. In fact, with the confining potential
obtained in $\left( \ref{confm}\right) ,$ we do not even have a normalizable
probability distribution function, so it can not even define an ensemble. We
see, in this way, a clear example for the failure of the mean-field
approximation. It is interesting to summarize what we obtain for the three
symmetries: 
\begin{equation}
\begin{tabular}{|l|l|l|}
\hline
& $Exact$ & $Mean-field$ \\ \hline
$\beta =1$ & $(\frac{N+1}{2})\ln (1+x^{2})$ & $(\frac{N-1}{2})\ln (1+x^{2})$
\\ \hline
$\beta =2$ & $N\ln (1+x^{2})$ & $~\frac{N}{2}\ln (1+x^{2})$ \\ \hline
$\beta =4$ & $(2N-1)\ln (1+x^{2})~$ & $~(\frac{N}{2}+\frac{1}{4})\ln
(1+x^{2})$ \\ \hline
\end{tabular}
.
\end{equation}
Notice that the mean-field analysis is still able to give the right
analytical expression for the part in $x$ of the potential, and again hints
that we should expect a linear polynomial for the part in $N$. Nevertheless,
it completely fails to give the right values for the coefficients of the
polynomial.

The typical use of the Wigner integral is to show, for example, that a
parabolic confining potential give rise to the Wigner semi-circle law \cite
{Meh}. This is standard material, but not completely trivial. Now, the
reader might wonder what happens with some intermediate cases, like for
example a density of eigenvalues without compact support but coming from a
strongly confining potential. A natural choice is a Gaussian distribution
for the density of states. This is also an interesting problem for other
reasons: on one hand, since we know in detail the Cauchy case -that turns
out to be rather simple- then, the Gaussian case naturally comes to mind. On
the other hand, a Gaussian density of states has a long story in the theory
of random matrices and it is indeed physical . We have to compute: 
\begin{equation}
V(x^{\prime })=\int_{-\infty }^{\infty }e^{-x^{2}}\ln \left| x-x^{\prime
}\right| dx~.
\end{equation}

This is a non trivial computation, relevant in other fields as well
(including disordered systems \cite{Cri} and determinants of random
Schr\"{o}dinger operators {\cite{Kni}}). For example, it appears in {\cite
{Kni}}, where it is left unsolved. We carry out the computation in Appendix
B. The result is:

\begin{equation}
V(x)=x_{~~2}^{2}F_{2}(1,1;\frac{3}{2},2;-x^{2})~.
\end{equation}

Note that we need a rather complex expression for the confining potential in
order to have an ensemble with Gaussian density of states in the limit $%
N\rightarrow \infty $.

In spite of this complex solution -in great contrast to the simplicity of
the Cauchy case-, it would be of interest to study it in connection with the
two body random matrix ensembles. These are old ensembles, that have been
the subject of considerable revision and renewed interest (see \cite{Ben}
for a recent review).They are designed to include correlations among the
different matrix elements and one of their main features is that they show,
in a certain regime, a Gaussian density of states. In contrast, we have
obtained the confining potential that give us this density of states for
ensembles with the typical symmetries ($\beta =1,2,4$) of random matrix
theory.

To conclude this section, notice the different results given by the first
corrective term in Dyson equation depending on the nature of the density of
states: 
\begin{equation}
\rho (x)=C\exp (-\sigma x^{\mu })\rightarrow \ln \rho (x)=\ln C-\sigma
x^{\mu }~,
\end{equation}

\begin{equation}
\rho(x)=\frac{\lambda}{\pi}\frac{1}{\lambda^{2}+x^{2}}\rightarrow\ln
\rho(x)=\ln\frac{\lambda}{\pi}-\ln(\lambda^{2}+x^{2})~.
\end{equation}

The constant terms are dropped from the potential since they are reabsorbed
in the normalization constant. Observe that the parameter appears explicitly
multiplying the potential in the stretched exponential case (in this case,
the parameter is essentially the variance and it is directly related to $N$%
). In contrast, in the Cauchy case, the parameter \footnote{%
That in the Cauchy case is not the variance, since it is infinite. This is
usually referred as lack of characteristic scale.} drops out, and does not
enter into the expression of the potential. We see then a big difference
between the two cases and a complete and systematic study of all the
possible cases seems worth studying.

\section{Conclusions and Outlook}

We have shown with simple arguments how to arrive to random matrix ensembles
that are the matrix generalization, in the sense of random matrix theory, of
the scalar Levy distributions. In the particular case where an exact
solution is known (the Cauchy case), an explicit form for the ensemble can
be obtained. We have arrived to known and well-established results in the
mathematics and the physics literature {\cite{Brou,Hua,Witt,Berc}}. More
important, our intuitive method of derivation clearly shows some of the
particular features of these type of ensembles. It reveals that we need an
explicit and, at most, linear dependence of the confining potential with the
dimension of the matrix. Namely, the confining potential is so weak that we
are forced to this dependence.

Thus, a purely logarithmic term (as exactly appears in the Cauchy case, and
asymptotically in the others), yields an algebraic expression for the weight
function, that has to battle in equal conditions with the Vandermonde
determinant. It is precisely this point what makes this ensemble so special.
It can also be seen within the Coulomb gas picture, where the Vandermonde
determinant is a logarithmic two-body repulsive term and the weight function
is essentially a logarithmic (one-body) confining potential, that needs the
explicit dependence with $N$ to be able to compensate the repulsive term.
Then, it is rather natural that we get the same expression for the density
of eigenvalues for all $N$, since we have a modified confining potential for
each $N$. This is exact for the Lorentzian ensemble and also for the tails
of the distribution of the eigenvalues for the other Levy cases.

The known results on the correlation functions of the Lorentzian ensemble
are also noteworthy. In addition with the previous arguments, they lead to
the expectattion that all the Levy ensembles possess the universal behavior.

We have also seen, in the discussion on mean-field approaches, the
differences with the Gaussian ensembles and with the strongly confining
paradigm in general. In particular, the computation of the confining
potential that gives a Gaussian shape for the density of states in the limit 
$N\rightarrow \infty $ is also useful to illustrate this. In the Levy cases,
due to its power-like behavior we need an explicit modification of the
potential for each dimension of the matrix. This leads to a relative simple
solution for the density of states that is essentially insensitive to the
level repulsion (by construction). This is exact in the Lorentzian case, and
in great contrast with the Gaussian ensembles, where its shape goes from a
Gaussian for $N=1$ to a semicircle when increasing $N$. We have also studied
the inverse case: after all the influence of level repulsion among all the
eigenvalues (in the limit $N$ very large), what do we need to end up with a
Gaussian shape for the density of states ? The solution is a very complex
expression for the confining potential. This complexity is in contrast with
the simplicity of the Cauchy case. One interesting remark is that a
power-like weight function is a natural antagonist for a power-like
repulsion term.

\section{Acknowledgments}

The author is indebted to Professor Piet Brouwer for very useful information
and to Adan-Love Garriga for a careful reading of the document and crucial
logistic help. Very constructive comments by an anonymous referee are
acknowledged as well.

\newpage

\appendix

\section*{A. The stable random variables}

The fundamental importance of the normal distribution is due to the Central
Limit Theorem which is a consequence of the Bernoulli and de Moivre-Laplace
theorems, and the law of large numbers.

For our discussion it is important to have clear the following basic results:

According to Levy, a distribution $F$ is stable if and only if, for the two
positive constants $c_{1}$ and $c_{2}$, there exists a positive constant $c$
such that $X$ given by: 
\begin{equation}
c_{1}X_{1}+c_{2}X_{2}=cX~,
\end{equation}

is a random variable following the same distribution $F $, as the
independent, identically distributed random variables $X_{1} $ and $X_{2} $.
Alternatively, if:

\begin{equation}
\varphi(z)\equiv\left\langle e^{iXz}\right\rangle =\int_{-\infty}^{\infty
}e^{iXz}dF(X)~,
\end{equation}

denotes the characteristic function of the distribution $F$, then $F$ is
stable if and only if: 
\begin{equation}
\varphi (c_{1}z)\varphi (c_{2}z)=\varphi (cz)~.
\end{equation}

The most general definition can be found in {\cite{Gne}}. Let $%
X,X_{1},X_{2},...,X_{n}$ be iid random variables with a common distribution $%
F$. Then $F$ is called stable iff and only iff there exist constants $%
c_{n}>0 $ and $\gamma _{n}$ such that: 
\begin{equation}
Y_{n}\equiv \sum_{i}X_{i}=c_{n}X+\gamma _{n}~.
\end{equation}
Then, the characteristic function according to the definition above satisfy
the functional relation: 
\begin{equation}
\varphi ^{n}(z)=\varphi (c_{n}z)e^{i\gamma _{n}z}~,
\end{equation}
which can be solved exactly, and the result is:

\subparagraph*{Proposition 1}

\begin{equation}
\psi(z)=log\varphi(z)=i\gamma z-c\left| z\right| ^{\alpha}\left\{
1+i\beta^{\prime}\frac{z}{\left| z\right| }w(z,\alpha)\right\} ~,
\end{equation}
where $\alpha,\beta^{\prime},\gamma,c $ are constants ($\gamma$ is any real
number, $0<\alpha\leq2, $ $-1<\beta^{\prime}<1 $, and $c>0 $, and,

\begin{equation}
\omega (z,\alpha )=\left\{ 
\begin{array}{c}
\tan \frac{\pi \alpha }{2}\quad if\,\alpha \neq 1 \\ 
\frac{2}{\pi }\log \left| z\right| \quad if\,\alpha =1~.
\end{array}
\right\}
\end{equation}
$\alpha $ is called the Levy index or characteristic exponent. The limiting
case $\alpha =2$ corresponds to the Gaussian. For $\beta ^{\prime }=0$ the
distribution is symmetric. $\gamma $ translates the distribution, and $c$ is
a scaling factor for $X$. So, these last two parameters are not essential
and one can disregard them.

\subparagraph*{Proposition 2.}

The asymptotic behavior of a Levy stable distribution follows the inverse
power-law: 
\begin{equation}
f_{\alpha,\beta^{\prime}}(x)\sim\frac{A_{\alpha,\beta^{\prime}}}{\left|
x\right| ^{1+\alpha}},\quad\alpha<2~.
\end{equation}

\subparagraph*{Proposition 3.}

The analytic form of a stable law is given through the Fox function:

\begin{equation}
f_{\alpha,\beta^{\prime}}\left( x\right) =\epsilon H_{2,2}^{1,1}\left[ x~%
\Bigg | 
\begin{array}{cc}
(1-\epsilon,\epsilon) & (1-\gamma,\gamma) \\ 
(~0~,~1~) & (1-\gamma,\gamma)
\end{array}
\right] ~,
\end{equation}

for $\alpha<1$. 
\begin{equation}
f_{\alpha,\beta^{\prime}}\left( x^{-1}\right) =\epsilon
x^{2}H_{2,2}^{1,1}\left[ x~\Bigg | 
\begin{array}{cc}
(-1,1~) & (-\gamma,\gamma) \\ 
(-\epsilon,\epsilon~) & (-\gamma,\gamma)
\end{array}
\right] ~.
\end{equation}
for $\alpha>1$. With the abbreviations $\epsilon=1/\alpha$ and $\gamma
=(\alpha-\beta^{\prime})/2\alpha$.

\subparagraph*{Examples:}

For $\alpha=2$, then $\beta^{\prime}\equiv0$, and the stable density is
identical to the Gaussian distribution.

For $\alpha=1 $ and $\beta^{\prime}=0 $, the stable density is identical to
the Cauchy or Lorentz distribution: 
\begin{equation}
f_{1,0}(x)=\frac{\lambda}{\pi(\lambda^{2}+x^{2})}~.
\end{equation}

Further examples can be found in {\cite{Gne,Sch}}.

\appendix

\section*{B.\ Integral computation.}

We need to compute the following integral: 
\begin{equation}
V(t)=\int_{-\infty }^{\infty }\ln \left| x-t\right| e^{-x^{2}}dx~.
\end{equation}
These type of integrals represent a characteristic polynomial computation
(the characteristic polynomial associated to a certain continuous density of
states). Its relevance in different fields of mathematics and physics can be
appreciated by consulting {\cite{Cri,Kni}}.

First, we consider the Taylor series expansion:

\begin{equation}
V(t)=\sum_{k=0}^{\infty}\frac{V^{(k)}(0)}{k!}t^{k}~.
\end{equation}
and we make the following change of variables $x=y+t $, then we have:

\begin{equation}
V(t)=\int_{-\infty }^{\infty }\ln \left| y\right| e^{-(y+t)^{2}}dy~.
\end{equation}
and we consider: 
\begin{equation}
\frac{d^{k}}{dz^{k}}e^{-z^{2}}=(-1)^{k}e^{-z^{2}}H_{k}(z)~.
\end{equation}
Then the Taylor expansion looks: 
\begin{equation}
V(t)=\sum_{k=0}^{\infty }\frac{(-1)^{k}\widehat{V_{k}}(0)}{k!}t^{k}~,
\end{equation}
where $\widehat{V_{k}(}0)=\int_{-\infty }^{\infty }\ln \left| x\right|
e^{-x^{2}}H_{k}(x)dx$. Because of the symmetry of the Hermite polynomials,
we arrive at: 
\begin{equation}
V(t)=\sum_{k=0}^{\infty }\frac{(-1)^{k}\widehat{V}_{2k}(0)}{k!}t^{k}\qquad 
\widehat{V}_{2k}(0)=2\int_{0}^{\infty }H_{2k}(x)e^{-x^{2}}\ln x~dx~.
\end{equation}
The point is that last integral can be solved and gives: 
\begin{equation}
\widehat{V}_{2k}(0)=-\frac{\sqrt{\pi }}{2}(-1)^{k}2^{2k}\Gamma (k)~.
\end{equation}
We obtain then: 
\begin{equation}
V(t)=-\frac{\sqrt{\pi }}{2}\left( \gamma +2\ln 2+\sum_{r=1}^{\infty }\frac{%
(-1)^{k}2^{2k}\Gamma (k)}{\Gamma (2k+1)}t^{2k}\right) ~.
\end{equation}
Using the recurrence and duplication formulas for the gamma function we can
arrive at our final result: 
\begin{equation}
V(t)=-\frac{\sqrt{\pi }}{2}\left( \gamma +2ln2-2t_{~~2}^{2}F_{2}(1,1;\frac{3%
}{2},2;-t^{2})\right) ~.
\end{equation}

\end{document}